%Paper: hep-th/9512119
%From: miramont@fpaxp1.usc.es
%Date: Fri, 15 Dec 95 20:32:45 -0200

%%%%%%%%%%%%%%%%%%%%%%%%%%%%%%%%%%%%%%%%%%%%%%%%%%%%%%%%%%%%%%%%%%%%%
%
%
%                                      Andrea PASQUINUCCI, 1988
%              PANDA.TEX               S.I.S.S.A., Trieste, Italy
%                                      (Revised 1991, Princeton, USA)
%
%--------------------------------------------------------------------
%
%    These are TEX macros. They work with PLAIN TEX (the basis
%    version of TEX). The only problem can be with the double-page
%    format since it depends on the type of software and laserwriter
%    you use to print, so I cannot guarantee that the double-page
%    format will work properly. Double-page MUST be printed in
%    LANDSCAPE orientation. (You shouldn't have troubles with fonts;
%    if you do, please let me know.)
%
%--------------------------------------------------------------------
%
%                     INTERACTIVE SECTION
%
%--------------------------------------------------------------------
%
\def\standardrisposta{s }\def\reducedrisposta{r }
\def\mplarisposta{mpla }\def\zerorisposta{z }
\def\doublerisposta{d }\def\cartarisposta{e }\def\amsrisposta{y }
\newcount\ingrandimento \newcount\sinnota \newcount\dimnota
\newcount\unoduecol \newdimen\collhsize \newdimen\tothsize
\newdimen\fullhsize \newcount\controllorisposta \sinnota=1
\newskip\infralinea  \global\controllorisposta=0
\immediate\write16 { ********  Welcome to PANDA macros (Plain TeX,
AP, 1991) ******** }
%\immediate\write16 { You'll have to answer a few questions in
%lowercase.}
%\message{>  Do you want it in double-page (d), reduced (r)
%or standard format (s) ? }\read-1 to\risposta
%
%\message{>  Do you want it in USA A4 (u) or EUROPEAN A4
%(e) paper size ? }\read-1 to\srisposta
%
%\message{>  Do you have AMSFonts 2.0 (math) fonts (y/n) ? }
%\read-1 to\arisposta
%
%--------------------------------------------------------------------
%
%             END INTERACTIVE SECTION - PAGE FORMATTING
%
%--------------------------------------------------------------------
%       The following parameters define defaults to the interactive
%       session.  At the moment I have set EUROPEAN and MATH FONTS
%
\def\risposta{s }
\def\srisposta{e }
\def\arisposta{y }
\ifx\risposta\standardrisposta \ingrandimento=1200
\message {>> This will come out UNREDUCED << }
\dimnota=2 \unoduecol=1 \global\controllorisposta=1 \fi
\ifx\risposta\reducedrisposta \ingrandimento=1095 \dimnota=1
\unoduecol=1  \global\controllorisposta=1
\message {>> This will come out REDUCED << } \fi
\ifx\risposta\doublerisposta \ingrandimento=1000 \dimnota=2
\unoduecol=2

\message {>> You must print this in
LANDSCAPE orientation << } \global\controllorisposta=1 \fi
\ifx\risposta\mplarisposta \ingrandimento=1000 \dimnota=1
\message {>> Mod. Phys. Lett. A format << }
\unoduecol=1 \global\controllorisposta=1 \fi
\ifx\risposta\zerorisposta \ingrandimento=1000 \dimnota=2
\message {>> Zero Magnification format << }
\unoduecol=1 \global\controllorisposta=1 \fi
\ifnum\controllorisposta=0  \ingrandimento=1200
\message {>>> ERROR IN INPUT, I ASSUME STANDARD
UNREDUCED FORMAT <<< }  \dimnota=2 \unoduecol=1 \fi
\magnification=\ingrandimento
%
%--------------------------------------------------------------------
%
%                        PARAMETERS SETTING
%
%  You can modify these parameters at your will (and resposability)
%--------------------------------------------------------------------
%
\newdimen\eucolumnsize \newdimen\eudoublehsize \newdimen\eudoublevsize
\newdimen\uscolumnsize \newdimen\usdoublehsize \newdimen\usdoublevsize
\newdimen\eusinglehsize \newdimen\eusinglevsize \newdimen\ussinglehsize
\newskip\standardbaselineskip \newdimen\ussinglevsize
\newskip\reducedbaselineskip \newskip\doublebaselineskip
\eucolumnsize=12.0truecm    % column h-size for european doublepage
                            % (12.0treucm default)
\eudoublehsize=25.5truecm   % sheet h-size for european duoblepage
                            % (25.5treucm default)
\eudoublevsize=6.7truein    % sheet v-size for european doublepage
                            % (6.5treuin default  or 17truecm?)
\uscolumnsize=4.4truein     % column h-size for american doublepage
                            % (4.4treuin default)
\usdoublehsize=9.4truein    % sheet h-size for american duoblepage
                            % (9.4treuin default)
\usdoublevsize=6.8truein    % sheet v-size for american doublepage
                            % (6.8treuin default)
\eusinglehsize=6.5truein    % sheet h-size for european singlepage
                            % (6.5truein default)
\eusinglevsize=24truecm     % sheet v-size for european singlepage
                            % (24truecm default)
\ussinglehsize=6.5truein    % sheet h-size for american singlepage
                            % (6.5truein default)
\ussinglevsize=8.9truein    % sheet v-size for american singlepage
                            % (8.9truein default)
\standardbaselineskip=16pt plus.2pt  % baselineskip for standard
                                     % format (16pt default)
\reducedbaselineskip=14pt plus.2pt   % baselineskip for reduced
                                     % format (14pt default)
\doublebaselineskip=12pt plus.2pt    % baselineskip for doublepage
                                     % format (12pt default)
%
%  \Portoffset and \Landoffset define the horizontal and vertical
%  offsets respectively for portrait and landscape modes. Example:
%  \def\Portoffset{\voffset=.4truein\hoffset=.125truein}
%
\def\Portoffset{}
\def\Landoffset{\voffset=-.2truein}
\ifx\risposta\mplarisposta \def\Portoffset{\hoffset=1.8truecm} \fi
%
%  \Landspec defines the \special command that sets the printer
%  to landscape mode without need to specify it directly in the
%  TeX to postscript translator (the command is site dependent).
%  Example: \def\Landspec{\special{ps: landscape}}
%
\def\Landspec{}
\tolerance=10000
\parskip=0pt plus2pt  \leftskip=0pt \rightskip=0pt
%
%   Do not modify anything of what follows
%                       (unless you know what you are doing!)
%----------------------------------------------------------------------
%
\ifx\risposta\standardrisposta \infralinea=\standardbaselineskip \fi
\ifx\risposta\reducedrisposta  \infralinea=\reducedbaselineskip \fi
\ifx\risposta\doublerisposta   \infralinea=\doublebaselineskip \fi
\ifx\risposta\mplarisposta     \infralinea=13pt \fi
\ifx\risposta\zerorisposta     \infralinea=12pt plus.2pt\fi
\ifnum\controllorisposta=0    \infralinea=\standardbaselineskip \fi
\ifx\risposta\doublerisposta   \Landoffset \else \Portoffset \fi
\ifx\risposta\doublerisposta \ifx\srisposta\cartarisposta
\tothsize=\eudoublehsize \collhsize=\eucolumnsize
\vsize=\eudoublevsize  \else  \tothsize=\usdoublehsize
\collhsize=\uscolumnsize \vsize=\usdoublevsize \fi \else
\ifx\srisposta\cartarisposta \tothsize=\eusinglehsize
\vsize=\eusinglevsize \else  \tothsize=\ussinglehsize
\vsize=\ussinglevsize \fi \collhsize=4.4truein \fi
\ifx\risposta\mplarisposta \tothsize=5.0truein
\vsize=7.8truein \collhsize=4.4truein \fi
%
%--------------------------------------------------------------------
%
%                            FONTS
%
%--------------------------------------------------------------------
%
\newcount\contaeuler \newcount\contacyrill \newcount\contaams
\font\ninerm=cmr9  \font\eightrm=cmr8  \font\sixrm=cmr6
\font\ninei=cmmi9  \font\eighti=cmmi8  \font\sixi=cmmi6
\font\ninesy=cmsy9  \font\eightsy=cmsy8  \font\sixsy=cmsy6
\font\ninebf=cmbx9  \font\eightbf=cmbx8  \font\sixbf=cmbx6
\font\ninett=cmtt9  \font\eighttt=cmtt8  \font\nineit=cmti9
\font\eightit=cmti8 \font\ninesl=cmsl9  \font\eightsl=cmsl8
\skewchar\ninei='177 \skewchar\eighti='177 \skewchar\sixi='177
\skewchar\ninesy='60 \skewchar\eightsy='60 \skewchar\sixsy='60
\hyphenchar\ninett=-1 \hyphenchar\eighttt=-1 \hyphenchar\tentt=-1
%
                 % math italic bold \bfmath
\font\tencmmib=cmmib10  \newfam\cmmibfam  \skewchar\tencmmib='177
                  % math bold (cal) symbols
\font\tencmbsy=cmbsy10  \newfam\cmbsyfam  \skewchar\tencmbsy='60
\def\scaps{\cmcsc}                 % small caps (uppercase)
\font\tencmcsc=cmcsc10  \newfam\cmcscfam
\ifnum\ingrandimento=1095

\font\capsone=cmcsc10 at 10.95pt \font\capstwo=cmcsc10 at 13.145pt

\else

\font\capsone=cmcsc10 at 12pt \font\capstwo=cmcsc10 at 14.4pt
\fi

\def\ttaarr{\bf}		% chapter titles' font
\def\ppaarr{\sl}		% section titles' font

%
     % inch-high caps (enormous)
%
%   AMS fonts (this works only if you have at least the 2.0
%              version of AMSFonts, otherwise say no)
%
\newfam\eufmfam \newfam\msamfam \newfam\msbmfam \newfam\eufbfam
\def\Loadeulerfonts{\global\contaeuler=1 \ifx\arisposta\amsrisposta
\font\teneufm=eufm10              %  \eufm   Gothic (or Euler)
\font\eighteufm=eufm8 \font\nineeufm=eufm9 \font\sixeufm=eufm6
\font\seveneufm=eufm7  \font\fiveeufm=eufm5
\font\teneufb=eufb10              %  \eufb   Bold Gothic (or Euler)
\font\eighteufb=eufb8 \font\nineeufb=eufb9 \font\sixeufb=eufb6
\font\seveneufb=eufb7  \font\fiveeufb=eufb5
\font\teneurm=eurm10              %  \eurm   Roman Gothic (or Euler)
\font\eighteurm=eurm8 \font\nineeurm=eurm9
\font\teneurb=eurb10              %  \eurb   Roman Bold Gothic
\font\eighteurb=eurb8 \font\nineeurb=eurb9
\font\teneusm=eusm10              %  \eusm   Slanted Capital Gothic
\font\eighteusm=eusm8 \font\nineeusm=eusm9
\font\teneusb=eusb10              %\eusb Slanted Capital Bold Gothic
\font\eighteusb=eusb8 \font\nineeusb=eusb9
\else \def\eufm{\tt} \def\eufb{\tt} \def\eurm{\tt} \def\eurb{\tt}
\def\eusm{\tt} \def\eusb{\tt}    \fi}
\def\loadeuler{\Loadeulerfonts\tenpoint}
\def\loadamsmath{\global\contaams=1 \ifx\arisposta\amsrisposta
\font\tenmsam=msam10 \font\ninemsam=msam9 \font\eightmsam=msam8
\font\sevenmsam=msam7 \font\sixmsam=msam6 \font\fivemsam=msam5
\font\tenmsbm=msbm10 \font\ninemsbm=msbm9 \font\eightmsbm=msbm8
\font\sevenmsbm=msbm7 \font\sixmsbm=msbm6 \font\fivemsbm=msbm5
\else \def\msbm{\bf} \fi \def\Bbb{\msbm} \def\symbl{\msam} \tenpoint}
\def\loadcyrill{\global\contacyrill=1 \ifx\arisposta\amsrisposta
\font\tenwncyr=wncyr10 \font\ninewncyr=wncyr9 \font\eightwncyr=wncyr8
\font\tenwncyb=wncyr10 \font\ninewncyb=wncyr9 \font\eightwncyb=wncyr8
\font\tenwncyi=wncyr10 \font\ninewncyi=wncyr9 \font\eightwncyi=wncyr8
\else \def\cyrill{\sl} \def\cyrilb{\sl} \def\cyrili{\sl} \fi\tenpoint}
\ifx\arisposta\amsrisposta
\font\sevenex=cmex7               %  reduced math symbols
\font\eightex=cmex8  \font\nineex=cmex9
\font\ninecmmib=cmmib9   \font\eightcmmib=cmmib8
\font\sevencmmib=cmmib7 \font\sixcmmib=cmmib6
\font\fivecmmib=cmmib5   \skewchar\ninecmmib='177
\skewchar\eightcmmib='177  \skewchar\sevencmmib='177
\skewchar\sixcmmib='177   \skewchar\fivecmmib='177
\font\ninecmbsy=cmbsy9    \font\eightcmbsy=cmbsy8
\font\sevencmbsy=cmbsy7  \font\sixcmbsy=cmbsy6
\font\fivecmbsy=cmbsy5   \skewchar\ninecmbsy='60
\skewchar\eightcmbsy='60  \skewchar\sevencmbsy='60
\skewchar\sixcmbsy='60    \skewchar\fivecmbsy='60
\font\ninecmcsc=cmcsc9    \font\eightcmcsc=cmcsc8     \else
\def\cmmib{\fam\cmmibfam\tencmmib}\textfont\cmmibfam=\tencmmib
\scriptfont\cmmibfam=\tencmmib \scriptscriptfont\cmmibfam=\tencmmib
\def\cmbsy{\fam\cmbsyfam\tencmbsy} \textfont\cmbsyfam=\tencmbsy
\scriptfont\cmbsyfam=\tencmbsy \scriptscriptfont\cmbsyfam=\tencmbsy
\scriptfont\cmcscfam=\tencmcsc \scriptscriptfont\cmcscfam=\tencmcsc
\def\cmcsc{\fam\cmcscfam\tencmcsc} \textfont\cmcscfam=\tencmcsc \fi
\catcode`@=11
\newskip\ttglue
\gdef\tenpoint{\def\rm{\fam0\tenrm}
  \textfont0=\tenrm \scriptfont0=\sevenrm \scriptscriptfont0=\fiverm
  \textfont1=\teni \scriptfont1=\seveni \scriptscriptfont1=\fivei
  \textfont2=\tensy \scriptfont2=\sevensy \scriptscriptfont2=\fivesy
  \textfont3=\tenex \scriptfont3=\tenex \scriptscriptfont3=\tenex
  \def\mcal{\fam2 \tensy}  \def\mmit{\fam1 \teni}
  \textfont\itfam=\tenit \def\it{\fam\itfam\tenit}
  \textfont\slfam=\tensl \def\sl{\fam\slfam\tensl}
  \textfont\ttfam=\tentt \scriptfont\ttfam=\eighttt
  \scriptscriptfont\ttfam=\eighttt  \def\tt{\fam\ttfam\tentt}
  \textfont\bffam=\tenbf \scriptfont\bffam=\sevenbf
  \scriptscriptfont\bffam=\fivebf \def\bf{\fam\bffam\tenbf}
     \ifx\arisposta\amsrisposta    \ifnum\contaeuler=1
  \textfont\eufmfam=\teneufm \scriptfont\eufmfam=\seveneufm
  \scriptscriptfont\eufmfam=\fiveeufm \def\eufm{\fam\eufmfam\teneufm}
  \textfont\eufbfam=\teneufb \scriptfont\eufbfam=\seveneufb
  \scriptscriptfont\eufbfam=\fiveeufb \def\eufb{\fam\eufbfam\teneufb}
  \def\eurm{\teneurm} \def\eurb{\teneurb} \def\eusm{\teneusm}
  \def\eusb{\teneusb}    \fi    \ifnum\contaams=1
  \textfont\msamfam=\tenmsam \scriptfont\msamfam=\sevenmsam
  \scriptscriptfont\msamfam=\fivemsam \def\msam{\fam\msamfam\tenmsam}
  \textfont\msbmfam=\tenmsbm \scriptfont\msbmfam=\sevenmsbm
  \scriptscriptfont\msbmfam=\fivemsbm \def\msbm{\fam\msbmfam\tenmsbm}
     \fi      \ifnum\contacyrill=1     \def\cyrill{\tenwncyr}
  \def\cyrilb{\tenwncyb}  \def\cyrili{\tenwncyi}         \fi
  \textfont3=\tenex \scriptfont3=\sevenex \scriptscriptfont3=\sevenex
  \def\cmmib{\fam\cmmibfam\tencmmib} \scriptfont\cmmibfam=\sevencmmib
  \textfont\cmmibfam=\tencmmib  \scriptscriptfont\cmmibfam=\fivecmmib
  \def\cmbsy{\fam\cmbsyfam\tencmbsy} \scriptfont\cmbsyfam=\sevencmbsy
  \textfont\cmbsyfam=\tencmbsy  \scriptscriptfont\cmbsyfam=\fivecmbsy
  \def\cmcsc{\fam\cmcscfam\tencmcsc} \scriptfont\cmcscfam=\eightcmcsc
  \textfont\cmcscfam=\tencmcsc \scriptscriptfont\cmcscfam=\eightcmcsc
     \fi            \tt \ttglue=.5em plus.25em minus.15em
  \normalbaselineskip=12pt
  \setbox\strutbox=\hbox{\vrule height8.5pt depth3.5pt width0pt}
  \let\sc=\eightrm \let\big=\tenbig   \normalbaselines
  \baselineskip=\infralinea  \rm}
\gdef\ninepoint{\def\rm{\fam0\ninerm}
  \textfont0=\ninerm \scriptfont0=\sixrm \scriptscriptfont0=\fiverm
  \textfont1=\ninei \scriptfont1=\sixi \scriptscriptfont1=\fivei
  \textfont2=\ninesy \scriptfont2=\sixsy \scriptscriptfont2=\fivesy
  \textfont3=\tenex \scriptfont3=\tenex \scriptscriptfont3=\tenex
  \def\mcal{\fam2 \ninesy}  \def\mmit{\fam1 \ninei}
  \textfont\itfam=\nineit \def\it{\fam\itfam\nineit}
  \textfont\slfam=\ninesl \def\sl{\fam\slfam\ninesl}
  \textfont\ttfam=\ninett \scriptfont\ttfam=\eighttt
  \scriptscriptfont\ttfam=\eighttt \def\tt{\fam\ttfam\ninett}
  \textfont\bffam=\ninebf \scriptfont\bffam=\sixbf
  \scriptscriptfont\bffam=\fivebf \def\bf{\fam\bffam\ninebf}
     \ifx\arisposta\amsrisposta  \ifnum\contaeuler=1
  \textfont\eufmfam=\nineeufm \scriptfont\eufmfam=\sixeufm
  \scriptscriptfont\eufmfam=\fiveeufm \def\eufm{\fam\eufmfam\nineeufm}
  \textfont\eufbfam=\nineeufb \scriptfont\eufbfam=\sixeufb
  \scriptscriptfont\eufbfam=\fiveeufb \def\eufb{\fam\eufbfam\nineeufb}
  \def\eurm{\nineeurm} \def\eurb{\nineeurb} \def\eusm{\nineeusm}
  \def\eusb{\nineeusb}     \fi   \ifnum\contaams=1
  \textfont\msamfam=\ninemsam \scriptfont\msamfam=\sixmsam
  \scriptscriptfont\msamfam=\fivemsam \def\msam{\fam\msamfam\ninemsam}
  \textfont\msbmfam=\ninemsbm \scriptfont\msbmfam=\sixmsbm
  \scriptscriptfont\msbmfam=\fivemsbm \def\msbm{\fam\msbmfam\ninemsbm}
     \fi       \ifnum\contacyrill=1     \def\cyrill{\ninewncyr}
  \def\cyrilb{\ninewncyb}  \def\cyrili{\ninewncyi}         \fi
  \textfont3=\nineex \scriptfont3=\sevenex \scriptscriptfont3=\sevenex
  \def\cmmib{\fam\cmmibfam\ninecmmib}  \textfont\cmmibfam=\ninecmmib
  \scriptfont\cmmibfam=\sixcmmib \scriptscriptfont\cmmibfam=\fivecmmib
  \def\cmbsy{\fam\cmbsyfam\ninecmbsy}  \textfont\cmbsyfam=\ninecmbsy
  \scriptfont\cmbsyfam=\sixcmbsy \scriptscriptfont\cmbsyfam=\fivecmbsy
  \def\cmcsc{\fam\cmcscfam\ninecmcsc} \scriptfont\cmcscfam=\eightcmcsc
  \textfont\cmcscfam=\ninecmcsc \scriptscriptfont\cmcscfam=\eightcmcsc
     \fi            \tt \ttglue=.5em plus.25em minus.15em
  \normalbaselineskip=11pt
  \setbox\strutbox=\hbox{\vrule height8pt depth3pt width0pt}
  \let\sc=\sevenrm \let\big=\ninebig \normalbaselines\rm}
\gdef\eightpoint{\def\rm{\fam0\eightrm}
  \textfont0=\eightrm \scriptfont0=\sixrm \scriptscriptfont0=\fiverm
  \textfont1=\eighti \scriptfont1=\sixi \scriptscriptfont1=\fivei
  \textfont2=\eightsy \scriptfont2=\sixsy \scriptscriptfont2=\fivesy
  \textfont3=\tenex \scriptfont3=\tenex \scriptscriptfont3=\tenex
  \def\mcal{\fam2 \eightsy}  \def\mmit{\fam1 \eighti}
  \textfont\itfam=\eightit \def\it{\fam\itfam\eightit}
  \textfont\slfam=\eightsl \def\sl{\fam\slfam\eightsl}
  \textfont\ttfam=\eighttt \scriptfont\ttfam=\eighttt
  \scriptscriptfont\ttfam=\eighttt \def\tt{\fam\ttfam\eighttt}
  \textfont\bffam=\eightbf \scriptfont\bffam=\sixbf
  \scriptscriptfont\bffam=\fivebf \def\bf{\fam\bffam\eightbf}
     \ifx\arisposta\amsrisposta   \ifnum\contaeuler=1
  \textfont\eufmfam=\eighteufm \scriptfont\eufmfam=\sixeufm
  \scriptscriptfont\eufmfam=\fiveeufm \def\eufm{\fam\eufmfam\eighteufm}
  \textfont\eufbfam=\eighteufb \scriptfont\eufbfam=\sixeufb
  \scriptscriptfont\eufbfam=\fiveeufb \def\eufb{\fam\eufbfam\eighteufb}
  \def\eurm{\eighteurm} \def\eurb{\eighteurb} \def\eusm{\eighteusm}
  \def\eusb{\eighteusb}       \fi    \ifnum\contaams=1
  \textfont\msamfam=\eightmsam \scriptfont\msamfam=\sixmsam
  \scriptscriptfont\msamfam=\fivemsam \def\msam{\fam\msamfam\eightmsam}
  \textfont\msbmfam=\eightmsbm \scriptfont\msbmfam=\sixmsbm
  \scriptscriptfont\msbmfam=\fivemsbm \def\msbm{\fam\msbmfam\eightmsbm}
     \fi       \ifnum\contacyrill=1     \def\cyrill{\eightwncyr}
  \def\cyrilb{\eightwncyb}  \def\cyrili{\eightwncyi}         \fi
  \textfont3=\eightex \scriptfont3=\sevenex \scriptscriptfont3=\sevenex
  \def\cmmib{\fam\cmmibfam\eightcmmib}  \textfont\cmmibfam=\eightcmmib
  \scriptfont\cmmibfam=\sixcmmib \scriptscriptfont\cmmibfam=\fivecmmib
  \def\cmbsy{\fam\cmbsyfam\eightcmbsy}  \textfont\cmbsyfam=\eightcmbsy
  \scriptfont\cmbsyfam=\sixcmbsy \scriptscriptfont\cmbsyfam=\fivecmbsy
  \def\cmcsc{\fam\cmcscfam\eightcmcsc} \scriptfont\cmcscfam=\eightcmcsc
  \textfont\cmcscfam=\eightcmcsc \scriptscriptfont\cmcscfam=\eightcmcsc
     \fi             \tt \ttglue=.5em plus.25em minus.15em
  \normalbaselineskip=9pt
  \setbox\strutbox=\hbox{\vrule height7pt depth2pt width0pt}
  \let\sc=\sixrm \let\big=\eightbig \normalbaselines\rm }
\gdef\tenbig#1{{\hbox{$\left#1\vbox to8.5pt{}\right.\n@space$}}}
\gdef\ninebig#1{{\hbox{$\textfont0=\tenrm\textfont2=\tensy
   \left#1\vbox to7.25pt{}\right.\n@space$}}}
\gdef\eightbig#1{{\hbox{$\textfont0=\ninerm\textfont2=\ninesy
   \left#1\vbox to6.5pt{}\right.\n@space$}}}
 %for 10-pt math in 9-pt territory
\def\alternativefont#1#2{\ifx\arisposta\amsrisposta \relax \else
\xdef#1{#2} \fi}
\global\contaeuler=0 \global\contacyrill=0 \global\contaams=0
%
%--------------------------------------------------------------------
%
%                            MACROS
%
%--------------------------------------------------------------------
%
\newbox\fotlinebb \newbox\hedlinebb \newbox\leftcolumn
\gdef\makeheadline{\vbox to 0pt{\vskip-22.5pt
     \fullline{\vbox to8.5pt{}\the\headline}\vss}\nointerlineskip}
\gdef\makehedlinebb{\vbox to 0pt{\vskip-22.5pt
     \fullline{\vbox to8.5pt{}\copy\hedlinebb\hfil
     \line{\hfill\the\headline\hfill}}\vss} \nointerlineskip}
\gdef\makefootline{\baselineskip=24pt \fullline{\the\footline}}
\gdef\makefotlinebb{\baselineskip=24pt
    \fullline{\copy\fotlinebb\hfil\line{\hfill\the\footline\hfill}}}
\gdef\doubleformat{\shipout\vbox{\Landspec\makehedlinebb
     \fullline{\box\leftcolumn\hfil\columnbox}\makefotlinebb}
     \advancepageno}
\gdef\columnbox{\leftline{\pagebody}}
\gdef\line#1{\hbox to\hsize{\hskip\leftskip#1\hskip\rightskip}}
\gdef\fullline#1{\hbox to\fullhsize{\hskip\leftskip{#1}%
\hskip\rightskip}}
\gdef\footnote#1{\let\@sf=\empty
         \ifhmode\edef\#sf{\spacefactor=\the\spacefactor}\/\fi
         #1\@sf\vfootnote{#1}}
\gdef\vfootnote#1{\insert\footins\bgroup
         \ifnum\dimnota=1  \eightpoint\fi
         \ifnum\dimnota=2  \ninepoint\fi
         \ifnum\dimnota=0  \tenpoint\fi
         \interlinepenalty=\interfootnotelinepenalty
         \splittopskip=\ht\strutbox
         \splitmaxdepth=\dp\strutbox \floatingpenalty=20000
         \leftskip=\oldssposta \rightskip=\olddsposta
         \spaceskip=0pt \xspaceskip=0pt
         \ifnum\sinnota=0   \textindent{#1}\fi
         \ifnum\sinnota=1   \item{#1}\fi
         \footstrut\futurelet\next\fo@t}
\gdef\fo@t{\ifcat\bgroup\noexpand\next \let\next\f@@t
             \else\let\next\f@t\fi \next}
\gdef\f@@t{\bgroup\aftergroup\@foot\let\next}
\gdef\f@t#1{#1\@foot} \gdef\@foot{\strut\egroup}
\gdef\footstrut{\vbox to\splittopskip{}}
\skip\footins=\bigskipamount
\count\footins=1000  \dimen\footins=8in
\catcode`@=12
\tenpoint
\ifnum\unoduecol=1 \hsize=\tothsize   \fullhsize=\tothsize \fi
\ifnum\unoduecol=2 \hsize=\collhsize  \fullhsize=\tothsize \fi
\global\let\lrcol=L      \ifnum\unoduecol=1
\output{\plainoutput{\ifnum\tipbnota=2 \clearnmbnota\fi}} \fi
\ifnum\unoduecol=2 \output{\if L\lrcol
     \global\setbox\leftcolumn=\columnbox
     \global\setbox\fotlinebb=\line{\hfill\the\footline\hfill}
     \global\setbox\hedlinebb=\line{\hfill\the\headline\hfill}
     \advancepageno  \global\let\lrcol=R
     \else  \doubleformat \global\let\lrcol=L \fi
     \ifnum\outputpenalty>-20000 \else\dosupereject\fi
     \ifnum\tipbnota=2\clearnmbnota\fi }\fi
\def\ifdoublepage{\ifnum\unoduecol=2 }
\gdef\yespagenumbers{\footline={\hss\tenrm\folio\hss}}
\gdef\ciao{ \ifnum\fdefcontre=1 \endfdef\fi
     \par\vfill\supereject \ifnum\unoduecol=2
     \if R\lrcol  \headline={}\nopagenumbers\null\vfill\eject
     \fi\fi \end}

\newskip\olddsposta \newskip\oldssposta
\global\oldssposta=\leftskip \global\olddsposta=\rightskip

\def\filldots{\leaders\hbox to 1em{\hss.\hss}\hfill}
\def\inquadrb#1 {\vbox {\hrule  \hbox{\vrule \vbox {\vskip .2cm
    \hbox {\ #1\ } \vskip .2cm } \vrule  }  \hrule} }
 \def\newline{\hfil\break}
\def\jump{\vskip\baselineskip} \newskip\iinnffrr
\def\sjump{\iinnffrr=\baselineskip
          \divide\iinnffrr by 2 \vskip\iinnffrr}
\def\bjump{\vskip\baselineskip \vskip\baselineskip}
\newcount\nmbnota  \def\clearnmbnota{\global\nmbnota=0}
\newcount\tipbnota \def\letterfootnote{\global\tipbnota=1}

\def\note#1{\global\advance\nmbnota by 1 \ifnum\tipbnota=1
    \footnote{$^{\rm\nttlett}$}{#1} \else {\ifnum\tipbnota=2
    \footnote{$^{\nttsymb}$}{#1}
    \else\footnote{$^{\the\nmbnota}$}{#1}\fi}\fi}
\def\nttlett{\ifcase\nmbnota \or a\or b\or c\or d\or e\or f\or
g\or h\or i\or j\or k\or l\or m\or n\or o\or p\or q\or r\or
s\or t\or u\or v\or w\or y\or x\or z\fi}
\def\nttsymb{\ifcase\nmbnota \or\dag\or\sharp\or\ddag\or\star\or
\natural\or\flat\or\clubsuit\or\diamondsuit\or\heartsuit
\or\spadesuit\fi}   \clearnmbnota
\def\numberfootnote{\global\tipbnota=0} \numberfootnote
\def\setnote#1{\expandafter\xdef\csname#1\endcsname{
\ifnum\tipbnota=1 {\rm\nttlett} \else {\ifnum\tipbnota=2
{\nttsymb} \else \the\nmbnota\fi}\fi} }
\newcount\nbmfig  \def\clearnbmfig{\global\nbmfig=0}
\gdef\figure{\global\advance\nbmfig by 1
      {\rm fig. \the\nbmfig}}   \clearnbmfig
\def\setfig#1{\expandafter\xdef\csname#1\endcsname{fig. \the\nbmfig}}

\newcount\frmcount \def\clearfrmcount{\global\frmcount=0}
\def\numero{\global\advance\frmcount by 1   \ifnum\indappcount=0
  {\ifnum\cpcount <1 {\hbox{\rm (\the\frmcount )}}  \else
  {\hbox{\rm (\the\cpcount .\the\frmcount )}} \fi}  \else
  {\hbox{\rm (\applett .\the\frmcount )}} \fi}
\def\nameformula#1{\global\advance\frmcount by 1%
\ifnum\draftnum=0  {\ifnum\indappcount=0%
{\ifnum\cpcount<1\xdef\spzzttrra{(\the\frmcount )}%
\else\xdef\spzzttrra{(\the\cpcount .\the\frmcount )}\fi}%
\else\xdef\spzzttrra{(\applett .\the\frmcount )}\fi}%
\else\xdef\spzzttrra{(#1)}\fi%
\expandafter\xdef\csname#1\endcsname{\spzzttrra}
\eqno \hbox{\rm\spzzttrra} $$}
\def\nfr{\nameformula}    
\def\nameali#1{\global\advance\frmcount by 1%
\ifnum\draftnum=0  {\ifnum\indappcount=0%
{\ifnum\cpcount<1\xdef\spzzttrra{(\the\frmcount )}%
\else\xdef\spzzttrra{(\the\cpcount .\the\frmcount )}\fi}%
\else\xdef\spzzttrra{(\applett .\the\frmcount )}\fi}%
\else\xdef\spzzttrra{(#1)}\fi%
\expandafter\xdef\csname#1\endcsname{\spzzttrra}
  \hbox{\rm\spzzttrra} }      \clearfrmcount
\newcount\cpcount \def\clearcpcount{\global\cpcount=0}
\newcount\subcpcount \def\clearsubcpcount{\global\subcpcount=0}
\newcount\appcount \def\clearappcount{\global\appcount=0}
\newcount\indappcount \def\clearindappcount{\indappcount=0}
\newcount\sottoparcount 

\def\applett{\ifcase\appcount  \or {A}\or {B}\or {C}\or
{D}\or {E}\or {F}\or {G}\or {H}\or {I}\or {J}\or {K}\or {L}\or
{M}\or {N}\or {O}\or {P}\or {Q}\or {R}\or {S}\or {T}\or {U}\or
{V}\or {W}\or {X}\or {Y}\or {Z}\fi    \ifnum\appcount<0
\immediate\write16 {Panda ERROR - Appendix: counter "appcount"
out of range}\fi  \ifnum\appcount>26  \immediate\write16 {Panda
ERROR - Appendix: counter "appcount" out of range}\fi}
\clearappcount  \clearindappcount \newcount\connttrre
\def\clearconnttrre{\global\connttrre=0} \newcount\countref
\def\clearcountref{\global\countref=0} \clearcountref
\def\chapter#1{\global\advance\cpcount by 1 \clearfrmcount
                 \goodbreak\null\vbox{\jump\nobreak
                 \clearsubcpcount\clearindappcount
                 \itemitem{\ttaarr\the\cpcount .\qquad}{\ttaarr #1}
                 \par\nobreak\jump\sjump}\nobreak}
\def\section#1{\global\advance\subcpcount by 1 \goodbreak\null
               \vbox{\sjump\nobreak\ifnum\indappcount=0
                 {\ifnum\cpcount=0 {\itemitem{\ppaarr
               .\the\subcpcount\quad\enskip\ }{\ppaarr #1}\par} \else
                 {\itemitem{\ppaarr\the\cpcount .\the\subcpcount\quad
                  \enskip\ }{\ppaarr #1} \par}  \fi}
                \else{\itemitem{\ppaarr\applett .\the\subcpcount\quad
                 \enskip\ }{\ppaarr #1}\par}\fi\nobreak\jump}\nobreak}
\clearsubcpcount
\def\appendix#1{\global\advance\appcount by 1 \clearfrmcount
                  \goodbreak\null\vbox{\jump\nobreak
                  \global\advance\indappcount by 1 \clearsubcpcount
          \itemitem{ }{\hskip-40pt\ttaarr #1}
%                  \itemitem{\ttaarr App.\applett\ }{\ttaarr #1}
             \nobreak\jump\sjump}\nobreak}
\clearappcount \clearindappcount
\def\references{\goodbreak\null\vbox{\jump\nobreak
   \noindent{\ttaarr References} \nobreak\jump\sjump}\nobreak}
%   \itemitem{}{\ttaarr References} \nobreak\jump\sjump}\nobreak}

\clearcpcount\clearcountref

\def\setchap#1{\ifnum\indappcount=0{\ifnum\subcpcount=0%
\xdef\spzzttrra{\the\cpcount}%
\else\xdef\spzzttrra{\the\cpcount .\the\subcpcount}\fi}
\else{\ifnum\subcpcount=0 \xdef\spzzttrra{\applett}%
\else\xdef\spzzttrra{\applett .\the\subcpcount}\fi}\fi
\expandafter\xdef\csname#1\endcsname{\spzzttrra}}
\newcount\draftnum \newcount\ppora   \newcount\ppminuti
\global\ppora=\time   \global\ppminuti=\time
\global\divide\ppora by 60  \draftnum=\ppora
\multiply\draftnum by 60    \global\advance\ppminuti by -\draftnum
\def\droggi{\number\day /\number\month /\number\year\ \the\ppora
:\the\ppminuti}     \global\draftnum=0
\def\draftcomment#1{\ifnum\draftnum=0 \relax \else
{\ {\bf ***}\ #1\ {\bf ***}\ }\fi} 
%
%     Maximum number of references = 200
%     boxes 50 -> 250 reserved for references
%
\catcode`@=11
\gdef\Ref#1{\expandafter\ifx\csname @rrxx@#1\endcsname\relax%
{\global\advance\countref by 1    \ifnum\countref>200
\immediate\write16 {Panda ERROR - Ref: maximum number of references
exceeded}  \expandafter\xdef\csname @rrxx@#1\endcsname{0}\else
\expandafter\xdef\csname @rrxx@#1\endcsname{\the\countref}\fi}\fi
\ifnum\draftnum=0 \csname @rrxx@#1\endcsname \else#1\fi}
\gdef\beginref{\ifnum\draftnum=0  \gdef\Rref{\fairef}
\gdef\endref{\scriviref} \else\relax\fi
\ifx\risposta\mplarisposta \ninepoint \fi
\parskip 2pt plus.2pt \baselineskip=12pt}
\def\Reflab#1{[#1]} \gdef\Rref#1#2{\item{\Reflab{#1}}{#2}}
\gdef\endref{\relax}  \newcount\conttemp
\gdef\fairef#1#2{\expandafter\ifx\csname @rrxx@#1\endcsname\relax
{\global\conttemp=0 \immediate\write16 {Panda ERROR - Ref: reference
[#1] undefined}} \else
{\global\conttemp=\csname @rrxx@#1\endcsname } \fi
\global\advance\conttemp by 50  \global\setbox\conttemp=\hbox{#2} }
\gdef\scriviref{\clearconnttrre\conttemp=50
\loop\ifnum\connttrre<\countref \advance\conttemp by 1
\advance\connttrre by 1
\item{\Reflab{\the\connttrre}}{\unhcopy\conttemp} \repeat}
\clearcountref \clearconnttrre
\catcode`@=12
\ifx\risposta\mplarisposta \def\Reflab#1{#1.} \letterfootnote \fi

\def\slashchar#1{\setbox0=\hbox{$#1$} \dimen0=\wd0
     \setbox1=\hbox{/} \dimen1=\wd1 \ifdim\dimen0>\dimen1
      \rlap{\hbox to \dimen0{\hfil/\hfil}} #1 \else
      \rlap{\hbox to \dimen1{\hfil$#1$\hfil}} / \fi}
\ifx\oldchi\undefined \let\oldchi=\chi
  \def\cchi{{\raise 1pt\hbox{$\oldchi$}}} \let\chi=\cchi \fi

\def\frac#1#2{{\textstyle{#1 \over #2}}}

\def\half{\ifinner {\scriptstyle {1 \over 2}}\else {1 \over 2} \fi}

\def\simge{\rlap{\raise 2pt \hbox{$>$}}{\lower 2pt \hbox{$\sim$}}}
\def\simle{\rlap{\raise 2pt \hbox{$<$}}{\lower 2pt \hbox{$\sim$}}}

\def\vbig#1#2{{\vbigd@men=#2\divide\vbigd@men by 2%
\hbox{$\left#1\vbox to \vbigd@men{}\right.\n@space$}}}

%
%--------------------------------------------------------------------
%
\newcount\fdefcontre \newcount\fdefcount \newcount\indcount
\newread\filefdef  \newread\fileftmp  \newwrite\filefdef
\newwrite\fileftmp     \def\strip#1*.A {#1}
\def\futuredef#1{\beginfdef
\expandafter\ifx\csname#1\endcsname\relax%
{\immediate\write\fileftmp {#1*.A}
\immediate\write16 {Panda Warning - fdef: macro "#1" on page
\the\pageno \space undefined}
\ifnum\draftnum=0 \expandafter\xdef\csname#1\endcsname{(?)}
\else \expandafter\xdef\csname#1\endcsname{(#1)} \fi
\global\advance\fdefcount by 1}\fi   \csname#1\endcsname}

\def\beginfdef{\ifnum\fdefcontre=0
\immediate\openin\filefdef \jobname.fdef
\immediate\openout\fileftmp \jobname.ftmp
\global\fdefcontre=1  \ifeof\filefdef \immediate\write16 {Panda
WARNING - fdef: file \jobname.fdef not found, run TeX again}
\else \immediate\read\filefdef to\spzzttrra
\global\advance\fdefcount by \spzzttrra
\indcount=0      \loop\ifnum\indcount<\fdefcount
\advance\indcount by 1   \immediate\read\filefdef to\spezttrra
\immediate\read\filefdef to\sppzttrra
\edef\spzzttrra{\expandafter\strip\spezttrra}
\immediate\write\fileftmp {\spzzttrra *.A}
\expandafter\xdef\csname\spzzttrra\endcsname{\sppzttrra}
\repeat \fi \immediate\closein\filefdef \fi}
\def\endfdef{\immediate\closeout\fileftmp   \ifnum\fdefcount>0
\immediate\openin\fileftmp \jobname.ftmp
\immediate\openout\filefdef \jobname.fdef
\immediate\write\filefdef {\the\fdefcount}   \indcount=0
\loop\ifnum\indcount<\fdefcount    \advance\indcount by 1
\immediate\read\fileftmp to\spezttrra
\edef\spzzttrra{\expandafter\strip\spezttrra}
\immediate\write\filefdef{\spzzttrra *.A}
\edef\spezttrra{\string{\csname\spzzttrra\endcsname\string}}
\iwritel\filefdef{\spezttrra}
\repeat  \immediate\closein\fileftmp \immediate\closeout\filefdef
\immediate\write16 {Panda Warning - fdef: Label(s) may have changed,
re-run TeX to get them right}\fi}
\def\iwritel#1#2{\newlinechar=-1
{\newlinechar=`\ \immediate\write#1{#2}}\newlinechar=-1}
\global\fdefcontre=0 \global\fdefcount=0 \global\indcount=0
%
%--------------------------------------------------------------------
%
\null
%
%--------------------------------------------------------------------
%
%                             THE    END
%
%--------------------------------------------------------------------
%
%\input panda
%\draftmode{Letter}
\loadamsmath
\loadeuler
\def\gg{{\widetilde g}}
\def\s{{\bf s}}
\def\sw{{{\bf s}_\omega}}
\def\Heis{{\cal H}[w]}
\def\ad{{\rm ad\;}}

\def\W{$\cal W$}
\pageno=0
\nopagenumbers{\baselineskip=12pt
\line{\hfill US-FT/31-95}
\line{\hfill\tt hep-th/9512119}
\line{\hfill December 1995}\bjump
\ifdoublepage \bjump\bjump\bjump\else\jump\vfill\fi
\centerline{\capstwo Deformed W--Algebras as Symmetries}
\sjump
\centerline{\capstwo of Generalized Integrable Hierarchies }
\bjump\bjump\jump
\centerline{{\scaps Carlos R.~Fern\'andez-Pousa\note{\tt
pousa@gaes.usc.es}}, {\scaps Manuel~V. Gallas\note{\tt
gallas@gaes.usc.es}},}\jump
\centerline{{\scaps J. Luis Miramontes\note{\tt
miramontes@gaes.usc.es}}, and {\scaps Joaqu\'\i n S\'anchez
Guill\'en\note{\tt joaquin@gaes.usc.es}}}
\jump\jump
\centerline{\sl Departamento de F\'\i sica de Part\'\i culas,}
\centerline{\sl Facultad de F\'\i sica,}
\centerline{\sl Universidad de Santiago,}
\centerline{\sl E-15706 Santiago de Compostela, Spain}
\bjump\bjump
\ifdoublepage
\vfill {\noindent
\line{December 1995\hfill}}
\eject\null\vfill\fi
\centerline{\capsone ABSTRACT}\jump

\noindent
A unified description of the relationship between the
Hamiltonian structure of a large class of integrable hierarchies of
equations and ${\cal W}$-algebras is discussed. The main result is an
explicit formula showing that the former can be understood as a
deformation of the latter.

\vfill
\ifdoublepage \else
\noindent
\line{December 1995\hfill}\fi
\eject}
\yespagenumbers\pageno=1
\footline={\hss\tenrm-- \folio\ --\hss}

Integrability and non-linear extensions of the conformal
algebra (\W-algebras) play an important role both in quantum
field theory and statistical mechanics. However, since they involve
non-linear structures, a systematic study is almost indispensable for
distinguishing their characteristics from the technicalities
involved in their mathematical description. The outstanding value of the
Drinfel'd and Sokolov's work~[\Ref{DS}] is precisely to provide a
systematic algebraic method for the construction of integrable
hierarchies of equations that unifies a large variety of previous
and dispersed results. More recently, and following~[\Ref{WIL}], a very
general approach that includes the original Drinfel'd-Sokolov
construction has been proposed in~[\Ref{GEN1},\Ref{GEN2}], and one of the
main features of the resulting generalized KdV-type hierarchies is
their early recognized relation with classical realizations of (deformed)
\W-algebras through their Hamiltonian structure. In this letter, we
discuss the general pattern of that relationship, worked out in detail
in~[\Ref{UNO},\Ref{DOS}], explaining in a simple and precise way the
main results and their implications. The crucial property to be explored
is the possibility of embedding the phase space of the
\W-algebras into the phase space of the hierarchies. Then, by
comparing the reduction processes involved in both constructions, it is
possible to investigate their equivalence, and, this way, to open
the possibility of obtaining new extensions of the conformal algebra.

In~[\Ref{GEN1},\Ref{GEN2}], generalized Drinfel'd--Sokolov (DS)
hierarchies of equations are constructed in terms of the matrix Lax
operator $L$ associated to the data $\{ \widetilde g, \Heis, {\bf
s}_\omega,{\bf s},\Lambda \}$. Here, $\widetilde g$ is the loop algebra
of a finite dimensional (complex) simple Lie algebra $g$, even though
the construction could be easily generalized to the case of reductive Lie
algebras too. $\bf s$ and ${\bf s_{\omega}}$ are two vectors of ${\rm
rank\/}(g)+1$ non-negative integers defining gradations of $\widetilde g$
such that ${\bf s} \preceq {\bf s}_\omega$ with respect to the partial
ordering introduced in [\Ref{GEN1},\Ref{GEN2}]; ${\bf s}_\omega$ also gives a
gradation of a Heisenberg subalgebra $\Heis$ of $\gg$. Finally,
$\Lambda$ is a constant element in $\Heis$ with
positive ${\bf s}_\omega$-grade $i$ that satisfies
$[\Lambda , {\widetilde g}_{<0}^{0}]\neq 0$\note{From now on,
superscripts and subscripts will indicate ${\bf s}_\omega$- and $\bf
s$-grades, respectively.}, which will be called the {\it non-degeneracy
condition}. It is worth mentioning that $\Lambda$ can be
equivalently characterized as a constant semisimple graded element in
$\widetilde g$ constrained by the latter condition. The corresponding Lax
operator
$ L = \partial_x + \Lambda(z) + q^{<i}_{\ge 0}(x,z)$ is defined in terms
of periodic currents (or potentials) of the form $\Lambda(z) +
q^{<i}_{\ge 0}(x,z)$, where $x \in  S^{1}$, and the
dependence on $z$, the affine parameter of the
loop algebra, is explicitly indicated. These Lax operators admit gauge
transformations preserving the form of the potential $q$,
$$
q\rightarrow  \widetilde q =\Phi \partial_{x} \Phi^{-1} +\Phi (\Lambda +
q^{<i}_{\ge 0}) \Phi^{-1} \,\,\, , \,\, \Phi \in G^{*}\>;
\nfr{gauge}
\noindent
they are generated by the gauge group $G^{*} = \exp(P)$ with $P=
\widetilde g_{0}^{<0}$, which is a nilpotent subalgebra of ${\widetilde
g}_0$. Gauge transformations preserve the infinite set of
commuting flows on $L$, which enables their restriction to gauge
equivalence classes. Moreover, the non-degeneracy condition ensures
the possibility of performing a~DS gauge-fixing thus leading to
gauge invariant currents which are polynomials on the original
currents.

The hierarchy also has a Hamiltonian description where the flows are
defined by means of a Poisson bracket and a set of infinite
Hamiltonians associated to the elements of ${\Heis}^{\geq0}$. Even more,
if $\bf  s$ is the homogeneous gradation, the hierarchy admits two
coordinated Poisson structures, but, in general, there is only one
that is usually called the ``second''. Its definition can be achieved
through a Poisson reduction procedure where the relevant algebraic
object is a classical $R$ matrix~[\Ref{SEMENOV}], {\it i.e.\/}, an
endomorphism of $\gg$ defined in terms of the gradation \s\ as
$R={1\over 2} [\Pi_{\ge 0} - \Pi_{<0}]$, with $\Pi_{\ge 0}$ and
$\Pi_{<0}$ being the projectors onto $\gg_{\geq0}$ and $\gg_{<0}$,
respectively. $R$ induces a different Lie algebraic structure on $\gg$,
whose Lie bracket will be denoted by $[\cdot,\cdot]_R$. The
corresponding Kirillov-Kostant bracket on the space of maps of $S^1$
onto $\gg$ is
$$
\eqalign{
\{\phi ,\psi \}[u]=&([d\phi ,d\psi]_R \bigm|
u)+\omega_R(d\phi\bigm|d\psi) \cr =&([\partial+u,d\phi_{\ge
0}]\bigm|d\psi_{\ge 0})-([u,d\phi_{<0}]\bigm|d\psi_{<0})\>,}
\nfr{hierro}
\noindent
where $(A\bigm|B) = \sum_{k \in {\Bbb Z}} \int_{S^1} dx \langle A_{k}(x)
, B_{-k}(x)\rangle$ is the generalization of the invariant Killing form
$\langle \cdot,\cdot \rangle$ of $g$ to the affine Lie algebra, $u(x,z) =
\sum_{k \in {\Bbb Z}} z^k u_k (x)$ is a generic $\gg$-current, and
$\omega_{R}$ is the associated $R$-cocycle~[\Ref{SEMENOV}]. This bracket,
restricted to the gauge invariant functionals of constrained
currents of the form $u(x,z)=\Lambda(z) +q^{<i}_{\ge 0}(x,z)$, is
precisely the ``second'' Poisson bracket~[\Ref{GEN2},\Ref{DOS}].

The infinite set of flow equations of the hierarchy is invariant
under a (global) scale transformation where the components of the
potential transform according to their ${\bf s}_{\omega}$-grades. This
result can be generalized to arbitrary conformal transformations which
are Poisson symmetries of the second Poisson bracket~[\Ref{GEN2}]; thus
suggesting a relationship between the second Poisson bracket algebra and
extended conformal algebras. Nevertheless, to establish rigourously
such relation one has to show that there exists a gauge invariant
energy-momentum tensor
$T_\epsilon [q]$ generating the conformal transformation,
$\delta_\epsilon q(x)=\{T_\epsilon, q(x)\}$. Actually, the generator for
the components of $q_0$ has already been obtained
in~[\Ref{NIGEL},\Ref{UNO}], but the existence of a generator
for those components lying on ${\widetilde g}_{>0}$ is
unclear. In particular, it is known that some of them are centres of
the second Poisson bracket and, hence, no energy-momentum
tensor can generate their conformal transformations.

Our purpose is to relate the second Poisson bracket
algebra with the \W-algebra associated to the finite reductive
subalgebra
${\widetilde g}_0$ and to its nilpotent element $\Lambda_0$, which
specifies an $sl(2,{\Bbb C})$ subalgebra of ${\widetilde
g}_0$~[\Ref{HAMRED},\Ref{FEHR}]. This \W-algebra is defined on the set of
invariant
${\widetilde g}_0$-currents with respect to the group of
transformations generated by some first class constraints; we will
show that it is a Poisson substructure of the second Poisson bracket
algebra. It is important to realize that the restriction of the
bracket~\hierro\ to the currents on ${\widetilde g}_0$ is just the
Kirillov-Poisson bracket associated to ${\widetilde g}_0$, which does
not involve the R-matrix at all. Then, the main difficulty in relating
the reduction procedures leading to the \W\ and the Poisson bracket
algebras is that the gauge transformations considered in the latter
generally mix the components
$q_0$ and $q_{>0}$ while those in the former are transformations of
$q_0$ only. However, we will be able to give a very precise account of
that relation by assuming an additional restriction
on the algebraic data defining the hierarchy. To be specific, we will
consider in more detail those cases where
$\Lambda +q_{\ge 0}^{<i} \in {\widetilde g}_0 \oplus {\widetilde
g}_1$; then the bracket of two gauge invariant functionals
of $\Lambda +q_{\ge 0}^{<i}$ reduces to
$$
\{\phi ,\psi \}[\Lambda +q]=([\partial+\Lambda_0 +q_{0}
^{<i},d\phi_{0}]\bigm|d\psi_{0})\>,
\nfr{buyo}
which is just the Kirillov-Poisson bracket on the reduced currents
$\Lambda_0 +q_{0} \in {\widetilde g}_0$.

In the Hamiltonian reduction approach, \W-algebras are defined by
means of the Dirac bracket associated to some second class
constraints on the set of currents associated to a finite (reductive)
Lie algebra~[\Ref{HAMRED}]. This Dirac bracket can be equivalently
described as the result of the reduction of the Kirillov-Kostant
bracket of the finite Lie algebra by first class constraints, {\it
i.e.\/}, of its Hamiltonian reduction~[\Ref{FEHR}]. In contrast, the
second Hamiltonian structure of the generalized Drinfel'd-Sokolov
hierarchies of~[\Ref{GEN1},\Ref{GEN2}] has a totally different origin.
It corresponds to a bracket on the set of gauge invariant functionals of
the Lax operator, which can be understood as the reduction of the
Kirillov-Kostant bracket associated to the R-dependent algebraic
structure defined now on a loop algebra~[\Ref{GEN2}]. To exhibit the
differences between both brackets, we start by
describing a non-standard reduction scheme that is general enough to
accomodate both reduction procedures; it has been introduced
by Feh\'er who names it ``hybrid reduction''~[\Ref{CONF}].

Let  $(M,\{ \cdot,\cdot \})$ be a Poisson manifold and
$\phi_1,\ldots,\phi_r$ a set of first class constraints~(FCC),
$\{\phi_i,\phi_j\}\bigm|_{\phi_k=0} = 0 $, such that their zero set
$N=\{~x\in M~\bigm|~\phi_k(x)=0~\}$ is an  embedded submanifold of
$M$. Under general conditions, the corresponding Hamiltonian vector
fields
$X_k(\cdot)=\{\phi_k,\cdot \}$ can be integrated to form a group
$G$ of transformations on $M$ preserving $N$; such transformations are
Hamiltonian by construction, {\it i.e.\/}, $\delta_k f=\{\phi_k,f\}$.
Let us assume that they are also Poisson transformations regular enough
to ensure  the existence of a space of orbits in
$N$, $N/G$, and of a convenient gauge slice $N/G \simeq \widehat M
\subset N$. Then, for any  function $f$ on $\widehat M$ there exists an
unique gauge invariant extension $\widehat  f$ onto $N$, which defines
the isomorphism of algebras
$\rho:C^\infty(\widehat M)
\rightarrow C^\infty _G(N)$, $\rho(f)={\widehat  f}$. Correspondingly,
this isomorphism induces the following Poisson structure on $\widehat M$:
$$ \{f,g\}^*=\rho ^{-1} ( \{\widehat f^\phi ,\widehat
g^\phi \} \bigm|_N )\>,
\nfr{lasa}
where $\widehat f^\phi$ and $\widehat g^\phi$ are arbitrary
extensions of $\widehat f$ and $\widehat g$ onto M, and the resulting
bracket is independent of the  choice of the extension as
a consequence of the first class character of the constraints. If the FCC do
not generate transformations on $N$, notice
that $\widehat M\equiv N$ is a Poisson submanifold. Therefore, since any subset
of constraints not generating transformations can be trivially imposed on
$M$, we will assume that all the FCC do generate a $r$-parameter group of
transformations on $N$.

Let us now suppose that the submanifold $\widehat M$ can be completely fixed
by an extended set of $2{r}$ constraints, $\{\psi_i\}=\{\phi_k\}\cup
\{\sigma_k\}$, $k=1,\ldots,r$, where the $\sigma_k$'s will be called
gauge fixing constraints. Then, the constraint matrix $\Delta_{ij}(x)=\{
\psi_i,
\psi_j \} (x) $ is non-degenerate on $\widehat M$, and, hence, it is
invertible. In this case, a convenient local
form of the Poisson structure is provided by the Dirac bracket
$$
\{f,g\}^*\equiv \{f,g\}^D= \{f^*,g
^*\} ~\bigm|_{\widehat M}  -\{f^*,\psi_i\}\Delta^{ij}\{\psi_j,g^*\}
\bigm|_{\widehat M}
\nfr{soler}
where $\Delta^{ij}(x)$ is the inverse of the constraint matrix,
and $f^*$ and  $g^*$ are arbitrary extensions of $f$ and $g$ onto $M$.
Since the gauge group is Hamiltonian by construction, this reduction
procedure is usually known as Hamiltonian reduction. Before proceeding,
let us point out that any centre $c$ of the Poisson structure is
invariant with respect to any Hamiltonian transformation, {\it
i.e.\/}, $\delta_k c=\{\phi_k,c\}=0$. Recall also that the
Dirac bracket can be generally defined in terms of any set of
constraints whose constraint matrix is nondegenerate (second
class constraints); the interpretation as a Hamiltonian reduction is
{\it a posteriori}.

However, to describe the second Poisson bracket algebra
we have to generalize the usual Hamiltonian reduction procedure.
Suppose that we enlarge the group of transformations from $G$ to
$G^*$,  $G \subset G^*$, such that $G^*$ induces another Poisson action
on $M$ preserving $N$. Assuming again regularity conditions for the
existence of $N/G^* \simeq \widehat M^*$, the same bracket {\lasa} now
defines a Poisson structure on $\widehat M^*$. When $G$ is trivial,
notice that this is just the Poisson reduction of $N = M$ by means of
the transformations generated by $G^*$. Nevertheless, the most
interesting case for our purposes occurs when the extended group of
transformations $G^*$ is not Hamiltonian, which implies that
the resulting bracket will not be of Dirac type in general.

Let us specialize the previous discussion to the reduction of the
currents involved in the construction of the second Poisson bracket
algebra:
$$
u(x,z) \in {\widetilde g} \longrightarrow \Lambda(z) \>+\> q^{<i}_{\ge
0}(x,z)\>.
$$
The set of currents $\Lambda + q_{\ge 0}^{<i}$ (the analogue of
the manifold $N$) can be viewed as the result of imposing the linear
constraints $ \phi_i[u(x)]= \Bigl\langle  \theta_i , u(x)-\Lambda
\Bigr\rangle = 0$ on the set of $\gg$-currents (corresponding  to
$M$) for any
$$
\theta_i ~\in ~ {\widetilde g}_{> 0} \oplus  {\widetilde g}_0^{\le -i}
\oplus {\widetilde g}_{<0}^{\le -i}\> \equiv \> \Gamma_{>0} \oplus
\Gamma_0 \oplus
\Gamma_{<0} \>.
\nfr{quique}
It can be easily checked that these constraints are first class, and that
only those associated to the nilpotent subalgebra $\Gamma_0$
generate transformations on
$N$~[\Ref{DOS}]. However, in general, the corresponding
Hamiltonian group of transformations is only a subgroup of the
group of gauge transformations generated by  $P ={\widetilde g}_0^{<
0}$, $\Gamma_0 = {\widetilde g}_0^{\le -i} \subseteq  P$.

Since the finite subalgebra ${\widetilde g}_0$ is graded by $\bf
s_\omega$, let $n$ be the highest $\bf s_\omega$-grade of the elements
of ${\widetilde g}_0$; for instance, when ${\bf s}=(1,0,\ldots,0)$ is
the homogeneous gradation\note{If $r= {\rm rank}(g)$, recall that the
gradations of the loop algebra $\widetilde g$ are specified by a set of
$r+1$ integers, ${\bf s}=({\bf s}^0,{\bf s}^1,\ldots,{\bf s}^r)$, and
that $N_{{\bf s}} = \sum_{j=0}^r k_j {\bf s}^j$, where $k_0, \ldots,
k_r$ are the (Kac) labels of the Dynkin diagram of $g$.} $n~=~N_{{\bf
s_\omega}}-{\bf s}_{\omega}^0$ is the $\sw$-grade of the
highest root step operator of $g$~[\Ref{UNO}]. Then, depending on the
values of
$n$ and
$i$, the
$\bf s_\omega$-grade of
$\Lambda$, the following cases can be distinguished:

(i) $i=1$, which means that $\Gamma_0=P$. Then, the group
of transformations is fully generated by FCC and the
reduction is just an example of Hamiltonian reduction. Moreover, and as
a special feature of this particular case, the restriction of the
Poisson bracket~\hierro\ to $N$ is well defined and, hence, the
second Poisson bracket algebra can be equivalently understood as a
Poisson reduction by means of the group $G=\exp(P)$.

(ii) $1<i\le n$ indicates that $\{0\}\neq \Gamma_0
\subset P$, but $\Gamma_0 \not= P$. Consequently, the set of
gauge transformations is larger that those generated
by the FCC and, therefore, they will not be Hamiltonian in general.
This case is a particular example of hybrid reduction.

(iii) Finally, $i>n$ implies that $\Gamma_0 =\{0\}$. Therefore, since
the constraints do not generate any transformation, the
constrained manifold $N$ is again a Poisson submanifold and the
second Poisson bracket algebra follows from a Poisson reduction.
Moreover, in this case, the projection of $\Lambda$ onto ${\widetilde
g}_0$ vanishes, $\Lambda_0=0$.
\sjump

Within the standard Hamiltonian reduction
approach, \W-algebras are constructed as follows~[\Ref{HAMRED},\Ref{FEHR}]. Let
${\widetilde g}_0$ be a finite reductive Lie algebra and consider the set of
currents on ${\widetilde g}_0$ equipped with the usual Kirillov--Kostant
bracket. Then, there is a \W-algebra for each embedded
$sl(2,{\Bbb C})$ subalgebra of ${\widetilde g}_0$\note{By  an embedding
of
$sl(2,{\Bbb C})$ into a reductive Lie algebra $\gg_0$ we mean a direct
sum of embeddings into each simple ideal.}, which is given by the Dirac
bracket associated to the second class constraints leading to
constrained currents whose components are lowest weights in the
decomposition of ${\widetilde g}_0$ under the adjoint action of the
$sl(2,{\Bbb C})$ subalgebra. In a completely equivalent way,
this  Dirac bracket can be derived by means of a Hamiltonian
reduction, {\it i.e.\/}, the second class constraints can be
split into first class and gauge fixing constraints, an explicit
decomposition technically known as a ``halving''[\Ref{FEHR}].

Our method to compare the second Poisson bracket algebra of the
integrable hierarchies of~[\Ref{GEN1},\Ref{GEN2}] with \W-algebras is
the following. When $\Lambda_0\not=0$, it is possible
to choose the gauge slice for the gauge transformations generated by
$G^*=\exp (P)$ such that its components on ${\widetilde g}_0$ are
lowest weights in the  decomposition of ${\widetilde g}_0$ under the
adjoint action of the $sl(2,{\Bbb C})$ subalgebra specified by
$\Lambda_0$. This way, the set of generators of the \W-algebra can be
embedded into the set of gauge invariant currents, and the corresponding
Poisson structures can be compared.

{}From now on, we will assume that $\Lambda + q_{\ge 0}^{<i} \subset
{\widetilde g}_0 \oplus {\widetilde g}_1$, which, in particular, implies
that $\Lambda$ can be uniquely decomposed as $\Lambda=\Lambda_0
+\Lambda_1$. Let us first consider the case when $\Lambda_0
\neq 0$, which requires $i\leq n$. Then, $J_+ = \Lambda_0$ is a
nilpotent element characterizing a $sl(2,{\Bbb C})$ subalgebra of
${\widetilde g}_0$, $(J_+,J_0, J_-)$, which induces the direct sum
decomposition $P=\bar P \oplus P^*$ with $P^*~=~{\rm Ker}(\ad~ J_+)~ \cap ~P$
and $\bar P~\cap ~ {\rm Ker}(\ad ~ J_+)~=\{0\}$.

Since gauge transformations act independently on the components of the
currents in ${\widetilde g}_0$ and
${\widetilde g}_1$, the transformations generated by the
elements of  $\bar P$ can be used to gauge fix the
components in ${\widetilde g}_0$ , while the remaining $P^*$ fix those in
${\widetilde g}_1$. This way, the gauge slice $q^{\rm can}$ can be
chosen such that $q^{\rm can}\cap {\widetilde g}_0 = {\rm
Ker}(\ad J_-)$, {\it i.e.\/}, such that the components of $q^{\rm
can}$ in ${\widetilde g}_0$ are lowest weights~[\Ref{DOS}]. This gauge
fixing amounts to impose the linear constraints associated to certain
subspace $\theta\subset {\widetilde g}_{-1}$, on
${\widetilde g}_1$, and to
$\Gamma_0\oplus\Gamma_1\oplus\Gamma_2\subset {\widetilde g}_0$, with
$$
\eqalign{
\Gamma_0= &{\rm Im} (\ad ~ J_-) \cap {\widetilde g}^{\le -i}_0\>, \cr
\Gamma_1= &{\rm Im} (\ad ~ J_-) \cap {\widetilde g}^{>-i}_0 \cap
{\widetilde g}^{<0}_0\>, \quad{\rm and}\quad
\Gamma_2= {\rm Im} (\ad ~ J_-) \cap {\widetilde g}^{\ge 0}_0\>,\cr}
\nfr{alcorta}
on ${\widetilde g}_0$. It is important to notice that  the constraints
associated to
$\theta$ can be considered independently of the others, since they lead
just to a Poisson subalgebra~[\Ref{DOS}]. Therefore, the
reduction of the phase space of currents on $\widetilde g$ by the
gauge transformations generated by $G^*=\exp(P)$ can be viewed as a two
steps process
$$
u \rightarrow \Lambda~ +~q_0~ +q^{\rm can}_{1}\rightarrow \Lambda~  +~
q^{\rm can}_0+~ q^{\rm can}_1\>,
$$
where the first reduction is trivial, {\it i.e.\/}, it leads just to a
Poisson  subalgebra. According to our previous discussion, we recognize
the constraints associated to
$\Gamma_0$ as the only FCC generating transformations on the reduced
phase space; moreover, by identifying
$[J_+,\Gamma_0]$ with $\Gamma_2^\vee$, the dual space of $\Gamma_2$, the
constraints associated to $\Gamma_2$ are precisely the gauge-fixing
constraints of those associated to $\Gamma_0$. In contrast, the origin of
the constraints generated by $\Gamma_1$ is that $P-\Gamma_0 \not=\{0\}$,
and, therefore, they have to be considered only if $i>1$.

The set of linear constraints induced by
$\Gamma_0\oplus\Gamma_1\oplus\Gamma_2$ on ${\widetilde g}_0$ is precisely the
same set  involved in the Hamiltonian reduction construction of the \W-algebra
associated to the
$sl(2,{\Bbb C})$ subalgebra $\{J_+=\Lambda_0, J_0, J_-\}$ of ${\widetilde
g}_0$. Consequently, not only the corresponding  constraint matrix
$\Delta_{ij}$ is non-degenerate, but also its
inverse $\Delta^{ij}$ exists everywhere on the phase space and it depends
polynomially on the reduced currents. Even more, since $\Delta_{ij}$ always
admits a ``halving"~[\Ref{FEHR}], the group of transformations $\widetilde
G$ generated by the FCC that specify the halving has to be a subgroup of
our $G^*$, although their actions on the currents will be different (in
general, $\widetilde G$ does not transform $q_1$ while $G^*$ does).

Therefore, our first result is that the phase space of
the hierarchies is an extension of the phase space of the
\W-algebra associated to ${\widetilde g}_0$ and $J_+=\Lambda_0$. Then, one
can define two {\it a priori} different brackets on the gauge invariant
currents: the second Poisson bracket  corresponding to the
$G^*$-invariant extensions, and the Dirac bracket giving the \W-algebra.
With respect to the latter, the components of $q_1^{\rm can}$ are just
centres, as those of $q_1$ before the reduction, but, in general,
this will not be the case with respect to the former.

In more geometrical terms, the Dirac bracket gives the Poisson structure
resulting from the Hamiltonian reduction of the currents $\Lambda_0 +
q^{<i}_{0}$ plus some trivial components (centres) $q_1$ by means of
$\widetilde G$. Then, for generic functionals $\phi$, $\psi$ of
$\Lambda + q^{\rm can}$, the Dirac bracket is given by
$$
\eqalign{
\{\phi\>, \> \psi\}^D \> &= \> \{\phi^*\>,\> \psi^*\}
\bigm|_{\Lambda+q^{\rm can}} \cr
\noalign{\vskip 0.2cm}
& \quad- \> \sum_{i,j}\> \int_{S^1} \> dx\> dy\>
\{\phi^*\>, \>\gamma_i(x)\} \> \Delta^{i,j}(x,y)\>
\{\gamma_j(y)\> ,\> \psi^*\}\>\bigm|_{\Lambda+q^{\rm can}} \>, \cr}
\nfr{milla}
where $\phi^*$, $\psi ^*$ are arbitrary extensions onto ${\widetilde
g}$, and $\{\gamma_i\}$ is some basis for the  vector subspace $\Gamma_0
\oplus \Gamma_1 \oplus \Gamma_2$. The constraint matrix $\Delta_{ij}$ is
a local differential operator whose block form is
$$
\eqalign{
\Delta_{i,j}(x,y)\> [q^{\rm can}] \> &= \>
\Bigl\langle [\gamma_i, \gamma_j]\>,\> J_+ + q^{\rm can}_0(x)
\Bigr\rangle\> \delta(x-y) ~+~\Bigl\langle \gamma_i,
\gamma_j\Bigr\rangle\> \partial_x \delta(x-y) \cr
& = \> \bordermatrix{ & \Gamma_0 & \Gamma_1
& \Gamma_2 \cr
\Gamma_0 & 0 & 0 & * \cr
\Gamma_1 & 0 & B(x)\delta(x-y) & * \cr
\Gamma_2 & * & * & * \cr}\> , }
\nfr{michel}
where the $*$'s stand for some matrix differential operators whose form is
irrelevant in the following, and,
for $\gamma_i, \gamma_j \in \Gamma_1$, $B_{i,j}(x)=\langle
[\gamma_i, \gamma_j]\>,\> J_+ + q^{\rm can}_0(x) \rangle$ is a $q^{\rm
can}_0$-dependent matrix. Consequently, the block form of the inverse matrix
is just
$$
\Delta^{i,j}(x,y)\> [q^{\rm can}] \> = \> \bordermatrix{ & \Gamma_0 &
\Gamma_1 &
\Gamma_2\cr
\Gamma_0 &  * & * & * \cr
\Gamma_1 &  * & B^{-1}(x)\delta(x-y) & 0 \cr
\Gamma_2 &  * & 0 & 0 \cr} \>.
\nfr{redondo}

To relate the two Poisson structures, let us choose
$\phi^*= \widehat \phi$ and $ \psi ^* = \widehat \psi$ being the
$G^*$-gauge invariant extensions of $\phi$ and $\psi$;  gauge
invariance implies that
$$
0\>= \>\delta_S\widehat \psi \> =\> \{ \widehat \psi, \phi_S\}[q] \> + \>
\Bigl( (d\widehat \psi)_{-1}\>, \> [S(x), \Lambda_1 + q_1(x)] \Bigr)\>,
\nfr{luisenrique}
\noindent
with $\phi_S[q]=(S(x),q(x))$ for any $S(x) \in P$, and this identity
already exhibits that some gauge transformations are not Hamiltonian.
Now, using {\luisenrique} in {\milla}, and taking into account the
explicit form~\redondo\ of the inverse constraint matrix, one obtains
$$
\eqalignno{
\{\phi,  \psi \}[q^{\rm can}] \> & = \> \{ \widehat \phi, \widehat \psi\}^D
[q]\>
+\> {\cal C}(\widehat \phi, \widehat \psi)\> [q]\>, \qquad {\rm
where} \cr
\noalign{\vskip 0.3cm}
{\cal C}(\widehat \phi, \widehat \psi)\> [q] \>  & = \>
\sum_{\gamma_i,\gamma_j \in \Gamma_1} \> \int_{S^1} dx\>
\Bigl\langle\> (d\widehat \phi)_{-1}\bigm|_{\Lambda+q^{\rm can}}\>,
\>[\gamma_i, \Lambda_1 +q^{\rm can}_1(x)]
\Bigr\rangle \cr
\noalign{\vskip 0.1cm}
 &\qquad\quad B^{i,j}[ q^{\rm can}_0(x)]\> \Bigl\langle\> (d\widehat
\psi)_{-1}\bigm|_{\Lambda+q^{\rm can}}\>,
\>[\gamma_j, \Lambda_1 + q^{\rm can}_1(x)]\Bigr\rangle\>,
&\nameali{laudrup} \cr}
$$
and we have explicitly indicated that $B^{i,j}(x)$ depends only on
$q^{\rm can}_0$. This last equation is our main result; it
shows that the second Poisson bracket algebra of the integrable
hierarchies of~[\Ref{GEN1},\Ref{GEN2}] is the
\W-algebra corresponding to $\{\cdot,\cdot\}^D$ deformed by ${\cal
C}(\cdot, \cdot)$, which is antisymmetric and depends polynomially on the
components of $q^{\rm can}$.

The resulting form of the second Poisson bracket algebra is clarified
by splitting the set of generators in
$W_a(x)$'s and $B_a(x)$'s, associated to the components of
$q^{\rm can}$ with $\bf  s$-grade zero and one, respectively. Then,
since $B_a(x)$ is always a centre of the Dirac bracket $\{\cdot
,\cdot\}^{D}$ but not of the second Poisson bracket, one gets
$$
\eqalign{
& \{W_a(y), W_b(z)\}\> = \> \{W_a(y),
W_b(z)\}^D \> + \> {\cal C}\left( W_a(y), W_b(z)\right) \cr
& \{W_a(y), B_b(z)\}\> = \> {\cal C}\left(W_a(y),
B_b(z)\right) \cr
& \{B_a(y), B_b(z)\}\> = \> {\cal C}\left(B_a(y),
B_b(z)\right) \>. \cr}
\nfr{raul}

We have already anticipated that the phase
space of the integrable hierarchies of~[\Ref{GEN1},\Ref{GEN2}] includes
non-dynamical components (centres) that should not be considered as
actual degrees of freedom; their elimination amounts to a trivial
further reduction of the Hamiltonian structure. Since they are
functionals of the components of $q^{\rm can}$, setting them
to zero provides additional (polynomial) relations that allow one to
express certain components in terms of the others. To be
precise~[\Ref{GEN2},\Ref{UNO}], when $i>1$ there is a centre of
$\{\cdot ,\cdot \}^{*}$ for each $b$ in the set
$$
{\cal Z}^*\>= \> \biggl[ {\rm Ker\/}(\ad \Lambda)\cap
\gg_{1-i}(\sw)\biggr]\> \cup \> \biggl[ {\rm Cent\/}\bigl({\rm Ker\/}(\ad
\Lambda)\bigr) \cap \Bigl[ \bigoplus_{j=1-i}^{-1} \gg_j(\sw)
\Bigr]\biggr]\>.
\nfr{esnaider}
Then~[\Ref{DOS}], if $(b)_{-1} \not = 0 $ it is possible to express
some components of
$q^{\rm can}_{1}$ in terms of those of $q^{\rm can}_{0}$. In contrast,
when $(b)_{-1}  = 0$, which is only possible if ${\rm Ker}(\ad  \Lambda) \not=
\Heis$ and, hence, $\Lambda$ is not regular, the elimination of this centre
implies that some of the {\it a priori} generators of the
\W-algebra have to be expressed in terms of the others; this latter
possibility is quite suggestive from the point of view of looking for new
extensions of the conformal algebra.

In the following, we produce some examples to illustrate  the power of
eq.\laudrup:

{\bf (i)} $\Lambda_0 \neq 0$ and
$P^*=\{0\}$. This is equivalent to ${\rm Ker}(\ad ~\Lambda_0)\cap
P=\{0\}$, which is a stronger version of our non-degeneracy condition
that naturally arises in the context of the Hamiltonian reduction
approach to
\W-algebras~[\Ref{FEHR}]. In this case, the condition
$\Lambda+q\in {\widetilde g}_0\oplus {\widetilde g}_1$, assumed to derive
eq.\laudrup, is actually unnecessary, and one can obtain an analogous
formula  when, instead, $\Lambda+q\in {\widetilde g}_0\oplus
\cdots \oplus {\widetilde g}_p$ but, still, $\Lambda_0\not=0$. Then,
the generalization of~\laudrup\ is given by
$$
\eqalign{ {\cal C}(\widehat \phi, \widehat \psi)\> [q] \>
& = \> \sum_{\gamma_i,\gamma_j \in \Gamma_1} \> \int_{S^1} dx\>
\Bigl\langle\> (d\widehat \phi)_{<0}\bigm|_{\Lambda+q^{\rm can}}\>,
\>[\gamma_i,
\Lambda_{>0}+q^{\rm can}_{>0}(x)] \Bigr\rangle \cr
\noalign{\vskip 0.1cm}\>& \qquad\quad B^{i,j}[ q^{\rm can}_0(x)]\>
\Bigl\langle\> (d\widehat
\psi)_{<0}\bigm|_{\Lambda+q^{\rm can}}\>, \>[\gamma_j, \Lambda_{>0} +
q^{\rm can}_{>0}(x)]  \Bigr\rangle\>.}
\nfr{amavisca}

According to~[\Ref{UNO}], this stronger version of the non-degeneracy
condition ensures that the gauge fixing can be performed in terms of the
components of $q_0$ only. Then, the $W_a(x)$'s only depend on $q_0$ and,
hence, $(dW_a(x))_{<0}=0$. Correspondingly, the generators $B_a(x)$,
associated to the components of $q^{\rm can}_{>0}$, have the general form
$$
B_a(x) \>= \>  \Bigl\langle \beta_a\>, \> q_1(x) \> +\>
\bigl[S^{\rm can}[q_0(x)]\>, \> q_1(x)\bigr]\> + \> \cdots
\Bigr\rangle \>,
\nfr{zamorano}
where $S^{\rm can}[q_0(x)]$ is the $q_0$-dependent gauge transformation
taking an arbitrary potential to its canonical form on the gauge slice,
and $\beta_a$ is an arbitrary constant element of the
subspace of ${\widetilde g}_{<0}$  that is dual to $q_{>0}^{\rm can}$. Then,
using that $S^{\rm can}[q_0(x)]$ vanishes for $q_0(x) \in q_0^{\rm can}(x)$, it
follows that
$$
(dB_a(x))_{<0} ~\bigm|_{\Lambda + q^{\rm can}} =\beta_a\> \delta(x-y)\>.
$$
Taking all this into account, the second Poisson bracket algebra gets
decoupled as~[\Ref{DOS}]
$$
\eqalign{
& \{W_a(y), W_b(z)\}\> = \> \{W_a(y), W_b(z)\}^D \cr
& \{W_a(y), B_b(z)\}\> = \> 0 \cr
& \{B_a(y), B_b(z)\}\> = \> {\cal
C}\left(B_a(y), B_b(z)\right) \>, \cr}
$$
and, hence, the $W_a(x)$'s, {\it i.e.}, the functionals of  $q^{\rm
can}_0$, form a \W-algebra~[\Ref{UNO}].

At this point, it is possible to address one of the relevant questions
regarding the connection between \W-algebras and integrable soliton
equations: is any \W-algebra the Poisson bracket
algebra of some Hamiltonian integrable hierarchy of equations? So far, the
answer seems to be negative. In fact, in the particular case of the
integrable hierarchies associated to $g=A_n$ and ${\bf s}=(1,0,\ldots,0)$
(the homogeneous gradation) such that ${\widetilde g}_0 \equiv g= A_n$,
this problem has been considered in~[\Ref{UNO}] where the following
result is obtained:

\sjump
{\it Within the {\W-algebras} constructed in terms of $A_n$,
only those associated to the embeddings of {$A_{1}$} into {$A_{n}$}
labelled by partitions of the form}
$$
n+1 = k(a) + q(1) \quad or \quad n+1 = k(a+1) + k(a) + q(1)\>,
\quad{\rm for}\quad a,k,q \in {\Bbb Z}\geq0\>,
$$
{\it correspond to the second Poisson bracket algebra of some of the
integrable hierarchies of~[\Ref{GEN1},\Ref{GEN2}]. Then, they involve the
Heisenberg subalgebras $\Heis\subset \widetilde{A}_n$ associated to the
conjugacy classes of the Weyl group of $A_n$ specified by $[w]=[k(a), q(1)]$
and
$[w]=[k(2a+1), (q-k)(1)]$, and $\Lambda$ has $\sw$-grade $1$
and $2$, respectively (see theorem~3 of~[\Ref{UNO}] for details).}\sjump

\noindent
Since the class of hierarchies considered
in~[\Ref{GEN1},\Ref{GEN2}] is large enough to accommodate practically all
the generalizations of the Drinfel'd-Sokolov construction so far
considered, we find this result particularly relevant.

Another general feature observed in~[\Ref{UNO}] is
that some \W-algebras are associated to more than one integrable
hierarchy. For example,  the
\W-algebras specified by the partitions $n+1 = k(2) + q(1) $ are shared
as second Poisson bracket algebras by the hierarchies associated to the
conjugacy classes $[w] = [k(2),q(1)] $ and
$[w] = [k(3),(q-k)(1)]$.

Particular examples are provided by the fractional $[2N+1]^{(2)}$
generalized KdV $A_{2N}$ hierarchies\note {Following the terminology
of~[\Ref{BAK}], the fractional
$[N]^{(i)}$ generalized KdV
$A_{N-1}$ hierarchy is associated to the  principal Heisenberg subalgebra
$\Heis= {\cal H}[N]$ of $\widetilde{A}_{N-1}$, the homogeneous gradation
${\bf s}=(1,0,\ldots,0)$, and the principal gradation ${\bf s_{\omega}}=
(1,1,\ldots,1)$. Then, $1 < i < N$ is the principal grade of $\Lambda$.},
where the  second Poisson bracket algebra is just the \W-algebra
associated to the $sl(2,\Bbb C)$ subalgebra labelled by the partition
$2N+1 = (N+1) +(N)$,  which is nothing else than the fractional
${\cal W}_{N}^{(2)}$ algebra of~[\Ref{OTROS1}]; this constitutes a
generalization of the results of~[\Ref{BAK},\Ref{GEN2},\Ref{TJIN}].

{\bf (ii)} $\Lambda_0 \neq 0$ and ${\rm dim\/} (P^*)=1$. Then, since
$P^*$ is one dimensional, the gauge fixing involves a unique component of
$q_1(x)$ and, hence, $(dW_a(x))_{-1}$ is a function of $x$ taking values
on certain one-dimensional subspace of ${\widetilde g}_{-1}$.
Consequently, since ${{\cal C}\left( W_a(y), W_b(z)\right)}$
is antisymmetric, this term vanishes identically and the $W_a$ generators
again form a \W-subalgebra of the second Poisson
bracket algebra~[\Ref{DOS}]. Nevertheless, in general, ${\cal C}\left(
B_a(y), W_b(z)\right)\not=0$ and, hence, the $W_a(x)$'s and the
$B_a(x)$'s are not decoupled.

As an example of this second case, the fractional $[N]^{(3)}$
generalized KdV $A_{N-1}$ hierarchies have been discussed in~[\Ref{DOS}].
Then, the Poisson bracket algebra is just the  \W-algebra
associated to the $sl(2,\Bbb C)$ subalgebra labelled by the
partition $N = ({N+2\over3}) + ({N-1\over3}) + ({N-1\over3})$ for $N \in
1 + 3 \Bbb Z$,  and $ N = ({N+1\over3}) + ({N+1\over3}) +
({N-2\over3})$ if $N \in 2 + 3 \Bbb Z$. In particular, $[4]^{(3)}$
corresponds just to the ``$\>\W_{4}^{(3)}\>$'' algebra of~[\Ref{OTROS2}].

{\bf (iii) } $\Lambda_0 \neq 0$ and ${\rm dim\/}(P^*) >1$. This is the most
general and interesting case but, so far, a systematic analysis is still to
be done. In particular, it is not known if some of the
corresponding Poisson bracket algebras give rise to new extensions
of the conformal algebra, and, in fact, the mere existence of a generator
for the conformal symmetry has not been investigated yet. On the contrary,
the  only example discussed in~[\Ref{DOS}] leads again to a \W-algebra.
It is the fractional $[N]^{(N-1)}$ generalized KdV
$A_{N-1}$ hierarchy, and the restriction of its second Poisson
bracket algebra to the $W_{a}$'s is the \W-algebra associated to
the partition $N = 2 + 1+\cdots +1 $~[\Ref{DOS}] (see also~[\Ref{CONF}]).

Finally, let us briefly discuss the form of the second Poisson bracket
algebra when $\Lambda_0=0$; always assuming that $\Lambda + q\in
{\widetilde g}_0 \oplus {\widetilde g}_1$. Then, since
$[\Lambda,P]=[\Lambda_{1},{\widetilde g}^{<0}_0]\subset  {\widetilde
g}_1$, the gauge fixing involves only the components of $q_1$ and,
therefore, the gauge invariant currents associated to the components of
$q^{\rm can}_0$ are of the form
$$
\eqalign{
W_a(x)\equiv W_{\alpha_a}(x)=&\Bigl\langle\alpha_a,q_0^{\rm
can}(x)\Bigr\rangle=\Bigl\langle
\alpha_a, e^{S[q_1(x)]}(\partial+q_0(x))e^{-S[q_1(x)]}\Bigr\rangle\cr
=&\Bigl\langle\alpha_a,q_0(x)-\partial S[q_1(x)] +[S[q_1(x)],q_0(x)]+
\cdots \Bigr\rangle~~,}
$$
for any $\alpha_a\in {\widetilde g}_0$, which means that
$ (dW_{\alpha_a}(x))_0\bigm|_{\Lambda + q^{\rm can}}=\alpha_a \delta (x-y) $.
Consequently, and according to~\buyo, the restriction of the Poisson
bracket to these currents is
$$
\{W_{\alpha_a}(x),W_{\alpha_b}(y)\}=W_{[\alpha_a,\alpha_b]}(x) \delta(x-y)\>
-\> \langle\alpha_a , \alpha_b\rangle\> \partial_x \delta(x-y)\>,
$$
which is just the Kirillov-Kostant bracket associated to ${\widetilde
g}_0$, {\it i.e.}, an affine Kac--Moody algebra. On the contrary, this
result is not valid when $\Lambda+q \not\subset {\widetilde g}_0 \oplus
{\widetilde g}_1$ even if the condition $\Lambda_0=0$ is satisfied,
which was not noticed in~[\Ref{UNO}].

To sum up, we have compared two different Poisson structures that can be
defined on the phase space of the generalized Drinfel'd-Sokolov
hierarchies of~[\Ref{GEN1},\Ref{GEN2}]: the second Poisson bracket
structure corresponding to their Hamiltonian formalism, and the Dirac
bracket defining some \W-algebra by means of Hamiltonian reduction. Both
are Poisson structures on the same phase space, the difference being the
$R$-matrix origin of the Poisson structure of the hierarchy that
changes the Lie-algebraic character of $\widetilde g$ and, hence, the
Hamiltonian nature of the relevant group of transformations.
Eq.\laudrup\ follows precisely from the breakdown of the Hamiltonian
property for some of the gauge transformations involved in the
construction of the hierarchies, and gives the precise relation between
both Poisson structures. In particular, it shows that the second Poisson
bracket algebra is a deformation of some \W-algebra. Although we have
succeed in the explanation of practically all the (polynomial)
\W-structures appearing in the literature in the context of generalized
Drinfel'd-Sokolov hierarchies, a more detailed analysis of the second
Poisson bracket algebra is still to be done. In particular, it is still
unclear if the resulting class of Poisson structures always
provide extensions, eventually new, of the conformal algebra; an important
question that eq.\laudrup\ should help to answer.

\sjump
\centerline{\bf Acknowledgements}

\noindent
The research reported in this paper has been supported partially
by C.I.C.Y.T. (AEN93-0729) and D.G.I.C.Y.T. (PB90-0772).

\references

\beginref
\Rref{DS}{V.G.~Drinfel'd and V.V.~Sokolov, J.~Sov. Math. {\bf 30} (1985) 1975;
Soviet. Math. Dokl. {\bf 23} (1981) 457.}
\Rref{WIL}{G.W.~Wilson, Ergod. Theor. \& Dyn. Sist. {\bf 1} (1981) 361.}
\Rref{GEN1}{M.F.~de Groot, T.J.~Hollowood, and J.L.~Miramontes, Commun. Math.
Phys. {\bf 145} (1992) 57.}
\Rref{GEN2}{N.J.~Burroughs, M.F.~de Groot, T.J.~Hollowood, and
J.L.~Miramontes, Commun. Math. Phys. {\bf 153} (1993) 187; Phys. Lett. {\bf
B277} (1992) 89.}
\Rref{UNO}{C.R.~Fern\'andez Pousa, M.V.~Gallas, J.L.~Miramontes, and
J.~S\'anchez Guill\'en,  Ann. Phys. (N.Y.) {\bf 243} (1995) 372.}
\Rref{DOS}{C.R.~Fern\'andez Pousa and J.L.~Miramontes, {\sl The Hamilton
Structure of Soliton Equations and Deformed {\W}-algebras\/}, US-FT/23-95,
SWAT/95/83, hep-th/9508084.}
\Rref{SEMENOV}{M.A.~Semenov-Tian-Shanskii, Func. Anal. Appl. {\bf 17}
(1983) 259; \newline A.G.~Reyman and M.A.~Semenov-Tian-Shanskii, Phys. Lett.
{\bf A130} (1988) 456.}
\Rref{NIGEL}{N.J.~Burroughs, Nonlinearity {\bf 6} (1993) 583; Nucl. Phys. {\bf
B379} (1992) 340.}
\Rref{HAMRED}{J.~Balog, L.~Feh\'er, P.~Forg\'acs, L.~O'Raifeartaigh, and
A.~Wipf, Ann. Phys. (N.Y.) {\bf 203} (1990) 76;\newline
F.A.~Bais, T.~Tjin, and P.~van Driel, Nucl. Phys. {\bf B357}
(1991) 632;\newline
L.~Feh\'er, L.~O'Raifeartaigh, P.~Ruelle, I.~Tsutsui,
and A.~Wipf, Ann. Phys. (N.Y.) {\bf 213} (1992) 1; \newline
L.~Frappat, E.~Ragoucy and P.~Sorba, Commun. Math. Phys. {\bf
157} (1993) 499;\newline
P.~Bouwknegt and K.~Schoutens, Phys. Rep. {\bf 223} (1993)
183.}
\Rref{FEHR}{L.~Feh\'er, L.~O'Raifeartaigh, P.~Ruelle,
I.~Tsutsui, and A.~Wipf, Phys. Rep. {\bf 222} (1992) 1;\newline
L.~Feh\'er,
L.~O'Raifeartaigh, P.~Ruelle, and I.~Tsutsui,  Commun. Math. Phys. {\bf 162}
(1994) 399.}
\Rref{CONF} {L.~Feh\'er, {\sl KdV type systems and \W-algebras in the
Drinfeld-Sokolov approach\/}, talk given given at the Marseille Conference on
\W-algebra, July 1995, hep-th/ 9510001.}
\Rref{OTROS1}{A. Polyakov, Int. J. Mod. Phys. {\bf A5} (1990) 833;
\newline
M.~Bershadsky, Commun. Math. Phys. {\bf A5} (1991) 833;  \newline
L.~Feh\'er, L.~O'Raifeartaigh, P.~Ruelle, and I.~Tsutsui, Phys. Lett.
{\bf B283} (1992) 243.}
\Rref{OTROS2}{D.A.~Depireux and P.~Mathieu, Int. J. Mod. Phys. {\bf A7}
(1992) 6053.}
\Rref{BAK}{I.~Bakas and D.A.~Depireux, Mod. Phys. Lett. {\bf A6}
(1991) 1561, ERRATUM ibid. {\bf A6} (1191) 2351;
Int. J. Mod. Phys. {\bf A7} (1992) 1767.}
\Rref{TJIN}{P.~van Driel, Phys. Lett. {\bf 274B} (1991) 179.}
\endref

\end